\documentclass[a4paper,11pt]{article}
\usepackage{pos}

\title{ALICE : online-offline processing for Run 3}

\author*[a]{David Rohr for the ALICE Collaboration}

\affiliation[a]{CERN,\\
  Geneva, Switzerland}

\emailAdd{drohr@jwdt.org}

\abstract{
ALICE will increase the data-taking rate for Run 3 significantly to 50 kHz continuous readout of minimum bias Pb--Pb collisions.
The foreseen reconstruction strategy consists of 2 phases: a first synchronous online reconstruction stage during data-taking enabling detector calibration, and a posterior calibrated asynchronous reconstruction stage.
The main challenges include processing and compression of 50 times more events per second than in Run 2, sophisticated compression and removal of TPC data not use for physics, tracking of TPC data in continuous readout, the TPC space-charge distortion calibrations, and in general running more reconstruction steps online compared to Run 2.
ALICE will leverage GPUs to facilitate the synchronous processing with the available resources.
In order to achieve the best utilization of the computing farm, we plan to offload also several steps of the asynchronous reconstruction to the GPU.
This paper gives an overview of the important processing steps during synchronous and asynchronous reconstruction and of the required computing capabilities.
}

\FullConference{%
  The Eighth Annual Conference on Large Hadron Collider Physics-LHCP2020 \\
  25-30 May, 2020\\
  online}

\begin{document}
\maketitle

\section{ALICE data processing during Run 3}

ALICE \cite{bib:alice} is undergoing major upgrades during the LHC long shutdown 2 in order to increase the heavy ion data-taking rate to 50 kHz of minimum bias data in continuous readout.
This involves a large computing upgrade and a change of the online and offline processing paradigm \cite{bib:o2tdr}.
Instead of a classical online trigger and QA (Quality Assurance) processing followed by posterior offline data reconstruction, the processing is split into a synchronous and an asynchronous phase.
In addition, ALICE does away with both software and hardware triggers and instead stores all recorded collisions in compressed form.
The main purpose of the synchronous processing, besides the known QA tasks, is  data compression and detector calibration while there is beam in the LHC and the experiment is running.
The compressed raw data and and the required input for the calibration is stored to a disk buffer at CERN and it is mirrored to the Tier 0 and 1 centers.
A short postprocessing produces the final calibration outputs.
When ALICE is not recording data, the online farm will read the data from the disk buffer and produce the final and fully calibrated reconstruction result in the asynchronous phase.
The asynchronous workload will be split between the online processing farm and the Tier 0 and 1 centers.

The largest impact on the reconstruction comes from  the switch from trigger-based MWPC (Multi Wire Proportional Chamber) readout of the TPC (Time Projection Chamber) detector to continuous readout using GEMs (Gas Electron Multipliers).
The operation of the GEM TPC at a high collision rate of 50 kHz will yield a huge space charge which creates distortions to the drift electrons of up to 20 cm which poses unprecedented challenges for the calibration.
In parallel, the continuous readout with the missing a priori assignment of hits in the TPC to primary collisions prevents the a priory transformation of TPC hits from native coordinates (time, pad row, and pad) to spatial coordinates.
As a consequence, the track seeding operates using the time as coordinate instead of the spatial $z$ coordinate.
The third challenge related to the TPC, which is by far the largest contributor to raw data, is the data compression.
The TPC creates around 3.4 terabyte per second of raw data, which must be compressed by the synchronous processing to a maximum of 100 gigabyte per second to fit into the available storage.
This happens in multiple steps, employing zero suppression in the FPGA based readout unit boards, removal of TPC hits of tracks not used for physics, reduction of the hit property entropy using multiple techniques, and finally entropy compression using ANS encoding.
Since the calibration and the data compression rely on TPC tracks, the synchronous phase performs full TPC tracking in real time \cite{bib:ctd2019}.

The TPC tracking is the most computing-intense workload of the synchronous reconstruction.
ALICE employs GPUs to speed up the processing and fit into the compute capacity available in the online farm.
The TPC tracking algorithm has been derived from the Run 2 HLT (High Level Trigger) TPC tracking \cite{bib:hltpaper} and has been adopted to the Run 3 conditions, in particular to the continuous read out \cite{bib:chep2019}.
One main difference compared to Run 2 is that the tracking does not process individual events but time frames of around 10 ms of continuous data as a whole.
Due to the drift-nature of the TPC detector and the continuous readout, the individual collisions in the time frame cannot be disentangled before the tracking.
Consequently, the full time frame must fit in GPU memory.
This requires efficient usage of the GPU resources and in particular the reusage of memory for sequential steps of the tracking algorithm \cite{bib:chep2019}.

\section{Synchronous processing performance}

\begin{figure}[htb]
 \begin{minipage}[t]{0.49\textwidth}
  \centering
  \includegraphics[height=2in]{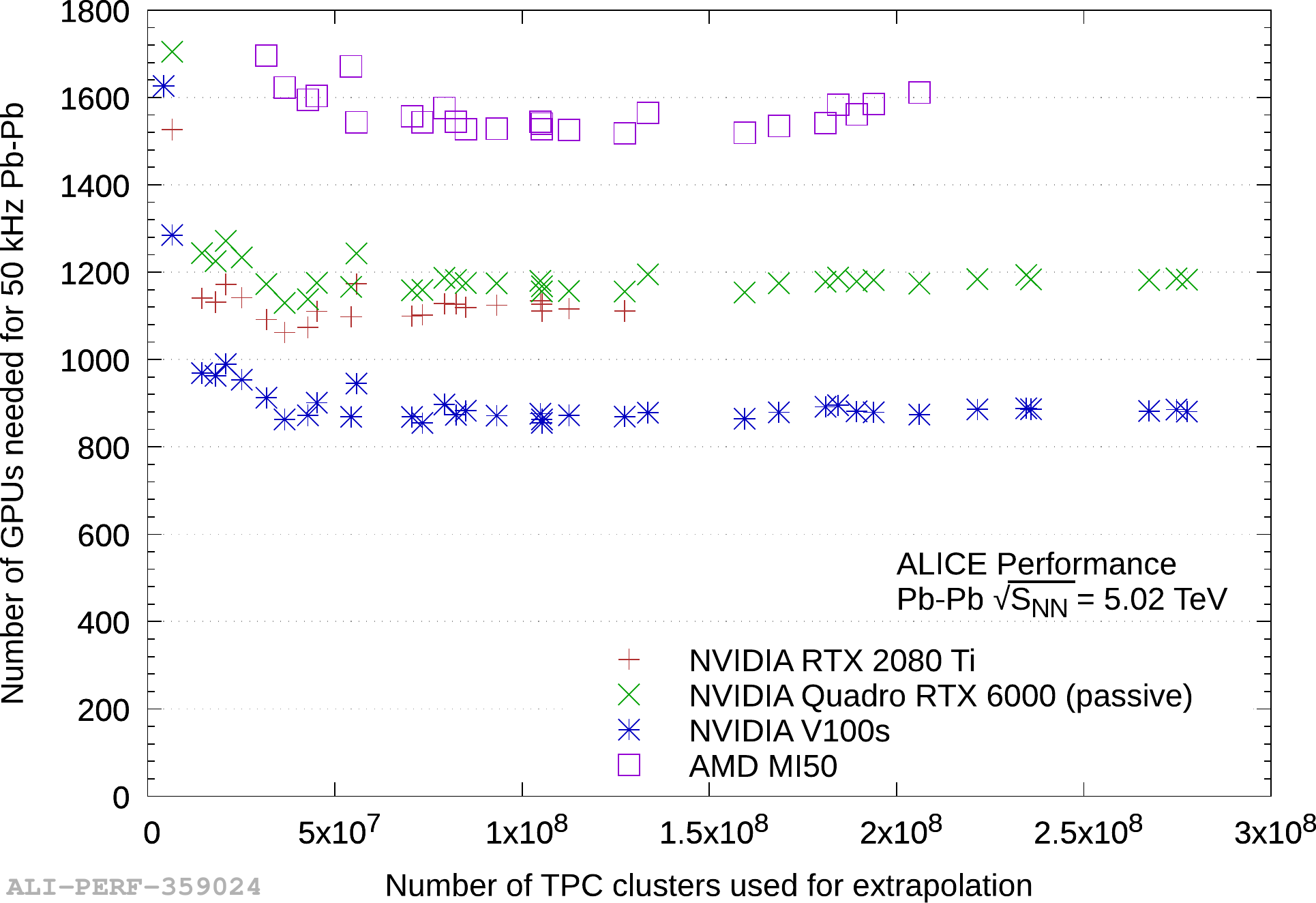}
  \caption{Number of GPUs needed for ALICE Run 3 synchronous processing.}
  \label{fig:ngpus}
 \end{minipage}
 \hfill
 \begin{minipage}[t]{0.49\textwidth}
  \centering
  \includegraphics[height=2in]{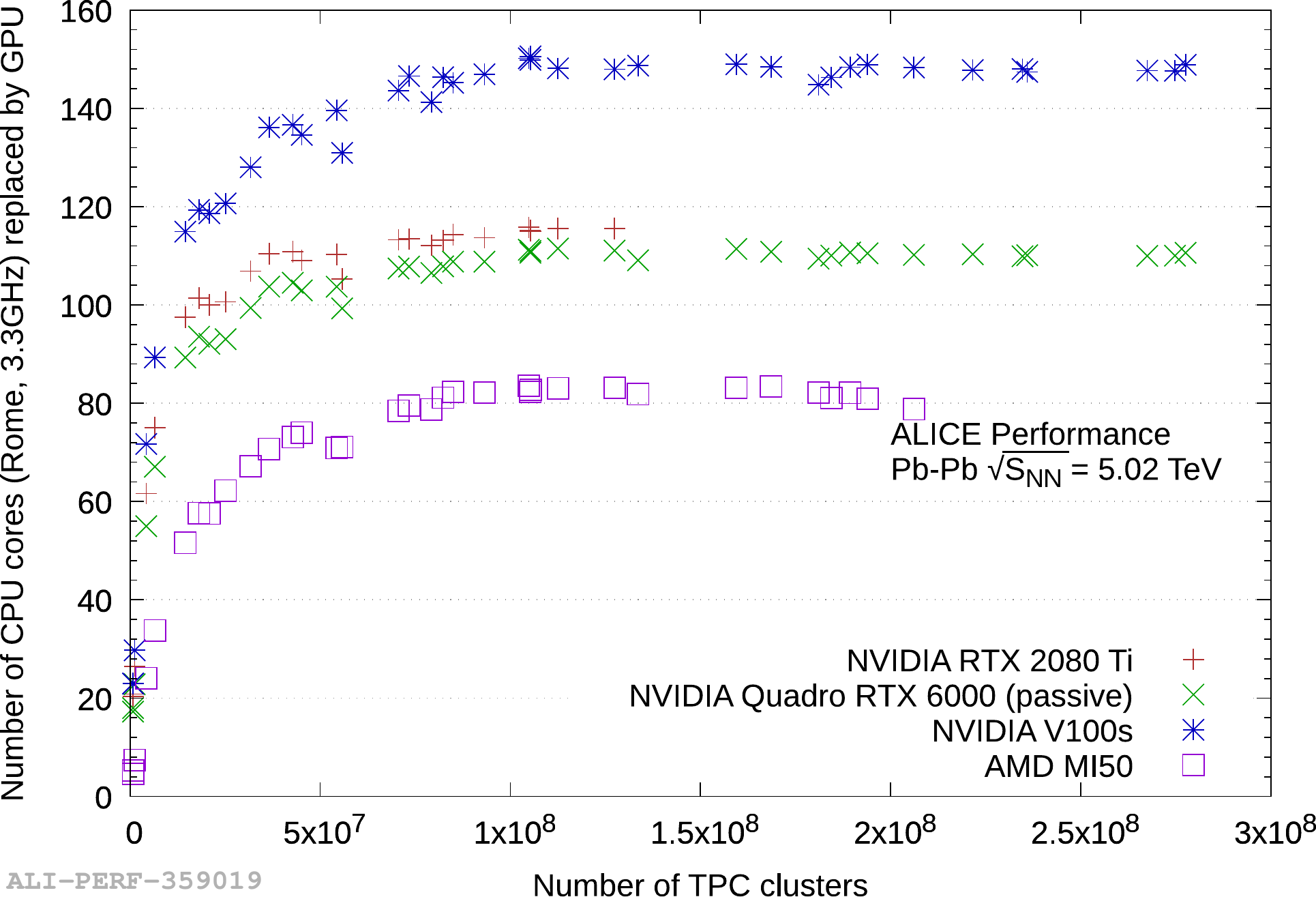}
  \caption{Speedup of GPUs in ALICE Run 3 synchronous processing versus 1 CPU core.}
  \label{fig:speedup}
 \end{minipage}
\end{figure}

The primary purpose of the online computing farm is the synchronous processing which consists to a large fraction of the TPC tracking.
Therefore, the hardware of the farm is chosen to maximize the TPC tracking performance within the given budget.
Since GPUs are much more powerful in this respect than classical processors, the GPUs will be the primary compute workhorses.
It is currently planned to acquire a farm of around 250 servers, each equipped with two 32-core processors and 8 GPUs.
The important number are the 2000 GPUs and the choice to place them in only 250 servers is a measure to reduce the infrastructure cost.
The required compute performance has been evaluated using a full system test of the reconstruction software processing simulated Monte-Carlo time frames converted to raw data.
The data is replayed from memory in a loop.
The test was running the full GPU reconstruction for the TPC and most of the CPU based reconstruction for other detectors, adding up to around 80\% of the required CPU capacity.
The CPU resources of the farm are well sufficient for the CPU workload measured during the full system test, including margin for the missing 20\% of reconstruction steps and for some additional CPU capacity required for operating the network and synchronizing the input data.
The resulting minimum number of GPUs based on this test is shown in Fig.~\ref{fig:ngpus} for various GPU models.
On top of this number, we require a margin of 20\% of GPU capacity which is needed to compensate for the fact that the online farm must not run at 100\% load and in order to enable future improvement of the tracking efficiency, which will require the fit of more tracks and more attached hits requiring additional computation time.
For all tested GPU models, 2000 units are sufficient.
The speedup of the GPU models compared to 1 core of an AMD Rome CPU running at 3.3 GHz is shown in Fig.~\ref{fig:speedup}.
We observe an almost linear weak scaling of the tracking performance on the CPU with multi-core CPUs up to 128 threads running on two 64-core Rome CPUs, thus the numbers can be scaled with the number of CPU cores to compare the GPU performance to a full CPU instead of a single core.

\section{Future plans for the asynchronous processing}

Even though the online computing farm is optimized for the synchronous processing, it is desirable to achieve a good utilization also during the asynchronous reconstruction.
The asynchronous reconstruction involves many more reconstruction steps of other detectors which are not necessary during the synchronous reconstruction.
In contrast, the TPC reconstruction part is faster than during the synchronous phase, because the asynchronous phase does not need to run the TPC clusterization and TPC compression parts and because the TPC input size is smaller after the synchronous phase has already discarded TPC hits not used for physics.
This yields a significant relative difference in the CPU and GPU usage comparing synchronous and asynchronous processing.
Naively, while the GPUs would be almost fully loaded in the synchronous phase, they would be mostly idling in the asynchronous phase.
Therefore, we aim to offload as many processing steps as possible to the GPU also in the asynchronous phase \cite{bib:ctd2020}.
This will improve the farm utilization and speed up the asynchronous processing in general.

A promising candidate for additional GPU offload is the full barrel tracking chain.
So far, in addition to the TPC tracking, a GPU version of the ITS (Inner Tracking System) exists and GPU tracking for the TRD (Transition Radiation Detector) is under development.
The aim is to have consecutive steps in the reconstruction chain on the GPU, such that the data can remain on the GPU all the time without intermediate transfer back and forth to the host.

\section{Conclusions}

ALICE will upgrade its computing towards the higher data rate in Run 3 and rely heavily on GPUs.
An on-site online computing farm will perform the synchronous processing which requires around 2000 GPUs.
Current work is ongoing to improve the GPU utilization during the asynchronous reconstruction, when the experiment is not running, to achieve best GPU utilization.


\begin{thebibliography}{99}

\bibitem{bib:alice}
{ALICE Collaboration},
``{The ALICE experiment at the CERN LHC}'',
J.~Inst. {\bf 3} S08002 (2008)

\bibitem{bib:o2tdr}
{ALICE Collaboration},
``Technical Design Report for the Upgrade of the Online-Offline Computing System'',
CERN-LHCC-2015-006, ALICE-TDR-019 (2015)

\bibitem{bib:ctd2019}
D.~Rohr for the {ALICE} Collaboration,
``{Global Track Reconstruction and Data Compression Strategy in ALICE for LHC Run 3}'',
{Proceedings of CTD2019} (2019)
arXiv:1910.12214

\bibitem{bib:hltpaper}
{ALICE Collaboration},
Real-time data processing in the ALICE High Level Trigger at the LHC,
CPC {\bf 242} 25 (2019),
arXiv:1812.08036

\bibitem{bib:chep2019}
D.~Rohr for the {ALICE} Collaboration,
``{GPU-based reconstruction and data compression at ALICE during LHC Run 3}'',
{Proceedings of CHEP 2019} (2019)
arXiv:2006.04158

\bibitem{bib:ctd2020}
D.~Rohr for the {ALICE} Collaboration,
``{Overview of online and offline reconstruction in ALICE for LHC Run 3}'',
{Proceedings of Connecting the Dots 2020} (2020)
arXiv:2009.07515

\end{thebibliography}
\end{document}